\documentclass[fleqn,usenatbib]{mnras}
\usepackage{newtxtext,newtxmath}
\usepackage[T1]{fontenc}
\DeclareRobustCommand{\VAN}[3]{#2}
\let\VANthebibliography\thebibliography
\def\thebibliography{\DeclareRobustCommand{\VAN}[3]{##3}\VANthebibliography}
\usepackage{graphicx}	
\usepackage{amsmath}	
\usepackage{caption}
\usepackage{subcaption}

\title[mass function non-universality]{Halo mass function in scale invariant models}

\author[Gavas et al.]{
Swati Gavas,$^{1}$\thanks{E-mail: swatigavas47@gmail.com}
Jasjeet Bagla,$^{1}$
Nishikanta Khandai$^{2,3}$
and Girish Kulkarni$^{4}$
\\
$^{1}$IISER Mohali, Knowledge city, Sector 81, SAS Nagar, Manauli PO 140306, Pujab, India\\
$^{2}$School of Physical Sciences, National Institute of Science Education and Research,
Jatni 752050, India \\
$^{3}$Homi Bhabha National Institute, Training School Complex,
Anushaktinagar, Mumbai 400094, India\\
$^{4}$Tata Institute of Fundamental Research, Homi Bhabha Road, Mumbai 400005, India
}

\date{Accepted XXX. Received YYY; in original form ZZZ}

\pubyear{2015}

\begin{document}
\label{firstpage}
\pagerange{\pageref{firstpage}--\pageref{lastpage}}
\maketitle

\begin{abstract}
Sheth-Tormen mass function has been widely used to quantify the abundance of dark matter halos.
It is a significant improvement over the Press-Schechter mass function as it uses ellipsoidal collapse in place of spherical collapse. Both of these mass functions can be written in a form that is universal, i.e., independent of cosmology and power spectrum when scaled in suitable variables.
However, cosmological simulations have shown that this universality is approximate.
In this paper, we investigate the power spectrum dependence of halo mass function through a suite of dark-matter-only N-body simulations of seven power-law models in an Einstein-de Sitter cosmology.
This choice of cosmology and a power-law power spectrum ensures the self-similar evolution of dark matter distribution, allowing us to isolate the power spectrum dependence of mass function.
We find that the mass function shows a clear non-universality.
We present fits for the parameters of the Sheth-Tormen mass function for a range of power-law power-spectrum indices. 
We find a mild evolution in the overall shape of the mass function with the epoch.
Finally, we extend our result to LCDM cosmology.  We show that the Sheth-Tormen mass function with parameter values derived from a matched power-law EdS cosmology provides a better fit to the LCDM mass function than the standard Sheth-Tormen mass function.  Our results indicate that an improved analytical theory is required to provide better fits to the mass function.
\end{abstract}



\section{Introduction}
\label{sec:intro}

Understanding the theoretical underpinnings of the large scale structure traced by galaxies and clusters of galaxies with precision has become essential with new state of the art observations.  
The halo model \citep{Peacock2000, Seljak2000,COORAY20021} assumes all mass in the Universe is contained in virialized dark matter halos.
It is formulated by quantifying the spatial distribution of halos and dark matter distribution within each halo.
The approach has been used successfully to make a wide variety of predictions \citep{Mandelbaum2005, Mead2015} and interpret observations \citep{Skibba2009, alam2020}.

The spatial distribution of halos is quantified using the excursion set formalism.  This formalism is also the basis of the theory of halo mass function.
These are key input components in various studies, e.g. Semi-analytical theories of galaxy formation \citep{1991ApJ...379...52W, somerville2015}, constraining cosmological parameters using galaxy cluster abundance \citep{Majumdar_2003, Allen2011, 2014A&A...571A..20P, to2021dark}, calculating halo merger rates \citep{Lacey1994, Cohn2001, Giocoli2008, Yacine2017, Saeed2021}, gravitational lensing \citep{bartelmann1997arc, Bartelmann_2010, Massey_2010} and constraining non-Gaussianity in the primordial power spectrum of matter perturbations \citep{Maggiore_2010, Dionysios2018, Shirasaki2021}.
\cite{Shimin2023} investigates the effects of variations in the halo mass function on the modelling of matter power spectra and the structure growth parameter $\sigma_8$ and finds that varying halo abundance may provide an alternative solution to the $\sigma_8$ tension.
Therefore understanding the theory of mass functions and its validity is critical for precision cosmology.  

The first detailed theory of the mass function of collapsed objects was given by \cite{press1974formation}(hereafter PS).
The model identifies spherical regions in initial linear Gaussian random fields representing peaks of the density field, which are dense enough to later collapses to form a non-linear structure.
The criterion for collapse is given by the mechanism of spherical collapse \citep{1972ApJ...176....1G}. 
\cite{1986ApJ...304...15B} provided a formal correspondence between the number density peaks in the matter density field and the number density of halos and provided a quantitative connection.  
Their theory underestimates mass function by a factor of two, as it does not consider that peaks that are part of bigger peaks should be excluded. 
Initial numerical results were good and in agreement with PS mass function theory \citep{Efstathiou1988, Carlberg1989, Lacey1994}.
\cite{Peacock1990} discussed conceptual problems with the PS mass function, such as not accounting mass in underdense regions, an assumption that formalism is independent of the choice of filter function and the lack of details over how overdense regions are related to bound objects.  
Several attempts were made to revisit the derivation of the mass function, but none of these could resolve the conceptual problems satisfactorily.  
Availability of larger and more accurate simulations exhibits clear ambiguity when compared with theory, specifically over-prediction of halos near characteristic mass($M_*$) and under-predicting at both ends. \citep{Monaco1997, Audit1998, Lee1999, Sheth1999}. 

An approach based on excursion set formalism was taken by \cite{1991ApJ...379..440B} to calculate the mass function.  
It uses Markovian random walks of mass elements and spherical up-crossing absorbing barrier to obtain the number density of halos.
These calculations are in Lagrangian space, which was then taken to Eulerian space by \cite{Mo1996} to address issues related to the clustering of halos.
This approach naturally resolves the discrepancy of factor two that arose in previous theories by using a top hat filter in the k-space \citep{Paranjape2012}.

The non-zero tidal field around collapsing region naturally leads to an ellipsoidal collapse rather than a spherical collapse.  
The shape of equal density contours around peaks in the density field is also not spherical.
These ideas were explored to modify PS mass function \citep{Monaco1997, Lee_1998, Sheth2001}.
\cite{2001MNRAS.323....1S} (hereafter ST) provided a functional form of ellipsoidal collapse barrier required in the excursion set formalism and derived mass function from it.

It is possible to develop the PS formalism while making no explicit reference to the cosmological model, the power spectrum of matter perturbations, and the epoch.
ST preserves this {\it universality} of the mass function.
These are simple theoretical predictions that do not encapsulate the intricacy of halo formation.
The accurate mass function can be obtained using cosmological N-Body simulations. 
Studies using simulations revealed a  $5-20 \%$ systematic deviation from theoretical predictions \citep{Jenkins2001, reed2006, Courtin2011, Diemer2020}.
Further, attempts have been made to examine the universality of mass function in various settings like cosmology\citep{Bhattacharya2011}, power spectra \citep{Ondaro2022}, epoch \citep{Watson2013}, mass definitions \citep{Despali2015}, mass interval \citep{reed2006}; showed a tendency to have a systematic trend in each case. 
This indicates that the complete picture requires us to go beyond the excursion set formalism.

Understanding the physical origin of non-universality is a tangled problem; it may be because of cosmology, power spectrum, or a combination of these.
Cosmology dependence is introduced by threshold density for collapse \citep{Barrow1993}.
The CDM class of power spectrum has a gradually varying slope; the spectral index $n(k)$ decreases with decreasing scale (increasing $k$).
As perturbations at smaller scales collapse earlier, the effective index of the power spectrum is small at an earlier time and increases towards a late time.
Therefore, variation in mass function and departures from a universal mass function may arise due to the changing slope of the power spectrum or cosmology, or both.
The only other option to work with fitting functions for each of these quantities derived from N-body simulations
(\citet{Jenkins2001, reed2003, Warren2006, reed2006, Tinker2008, Crocce2010, Bhattacharya2011, Courtin2011, Angulo2012, Watson2013, Bocquet2015, Despali2015, Comparat2017, McClintock_2019, Nishimichi_2019, Bocquet_2020,  Diemer2020, Seppi2021}).
The effect of baryons on mass function is seen in large cosmological hydrodynamic simulations \citep{Khandai2015}.

Since CDM spectra lack the simplicity of scale-free spectra, we approach the problem of non-universality differently.
We specifically look for the departures from non-universality in the mass function for scale-free power spectra of initial fluctuations with an Einstein-de Sitter(EdS) background to check if non-universal description can be attributed to a spectrum dependence.
Our choice of cosmology does not introduce any scale in the problem, and the threshold overdensity does not vary with time in any non-trivial manner.
Thus we can isolate the non-universality of mass function arising from the slope of the power spectrum.
We provide spectrum dependent fits for the parameters in the ST mass function and show that this allows us to better fit simulation data.

\cite{bagla2009mass} have explored this idea with a similar approach.  The key results of their work overlap with our findings, though as we shall show, the results presented here are demonstrated to be robust across many variations in the analysis. A similar analysis was also taken up by \citet{2022arXiv220802174E} who
see a similar departure from universality of the mass function. 

In section \ref{sec:formalism} we begin with the basic framework of mass function theory.
We discuss methodology in section \ref{sec:method}, where we describe our simulations.
In section \ref{sec:results} we present results from our data analysis.
We extend this work to LCDM models in section \ref{sec:lcdm}. 
Finally, we conclude our findings in section \ref{sec:conclusion}.

\section{Mass function: Theory}
\label{sec:formalism}

The form of equations which was thought to be universal \citep{Lacey1994} for PS and ST mass function is given by 
\begin{equation}
    f(\nu) =\nu \sqrt{\frac{2}{\pi}}  \exp(-\nu^2 /2) \text{, and}
    \label{eqn:ps}
\end{equation}
\begin{equation}
    f(\nu) = A(p) \nu \sqrt{\frac{2q}{\pi}} [1+(q\nu^2)^{-p}]  \exp(-q\nu^2 /2) \text{,}
	\label{eqn:st}
\end{equation}
where  $\nu = \delta_c /(D_+(z)\sigma(m))$. 
Here $\delta_c$ is the linearly extrapolated critical overdensity required for a spherically symmetric perturbation, above which the region collapses to form a virialized halo.
It generally has a weak dependence on cosmology, but its value is a constant for EdS cosmology. 
$D_+$ is the growth factor, which is proportional to the scale factor for EdS cosmology. 
$\sigma(m)$ is the variance in the initial density fluctuation field, linearly  extrapolated to the present epoch and smoothed with a top hat filter ($W(k,m)$) of scale R= $(3m/4\pi\bar{\rho})^{1/3}$.
It is calculated by convolving power spectra $P(k)$ with the filter function 
\begin{equation}
    \sigma^2(m) = \int_0^{\infty} \frac{dk}{k} \frac{k^3 P(k)}{2\pi^2} W^2(k,m) \text{.}
    \label{eqn:sgm}
\end{equation}

Both PS and ST formalism assume that all the mass in the Universe is present in the form of halos of some mass, i.e., $\int_0^{\infty} f(\nu)d\nu /\nu = 1$.
This provides the normalization for the mass functions.
\begin{equation}
    A(p) = \left[1+ \frac{2^{-p} \Gamma(0.5-p)}{\sqrt{\pi}}\right]^{-1}
    \label{eqn:Ap}
\end{equation}
where, $p \approx 0.3$ and $q \approx 0.75$ are the standard values proposed by ST.

These mass functions can be connected to the number density of halos of a given mass (per unit comoving volume).
\begin{equation}
    \frac{dn}{d \ln m} = \frac{\bar{\rho}}{m} \frac{d \ln \sigma^{-1}}{d \ln m} f(\nu)
    \label{eqn:dn}
\end{equation}
The dependence of the mass function on cosmology comes mainly through the growth factor $D_+$ and small dependence from threshold density $\delta_c$, and all the power spectrum dependence lies in mass variance $\sigma(m)$.
Both of these dependencies are absorbed in the definition of $\nu$.
Thus equations \ref{eqn:ps} and \ref{eqn:st} have a universal form.
The PS mass function has only one parameter, $\delta_c$, and this is fixed by theoretical considerations.
The ST mass function has two more free parameters $p$ and $q$. 
It is our aim to study whether these parameters are independent of the slope of the power spectrum or not. 
If the theory of mass function can be constructed in a truly universal fashion then these should be independent of the power spectrum. 

\section{Methodology}
\label{sec:method}

\subsection{Simulations}
\label{subsec:simulations}

\begin{table*}
	\begin{center}
	 \begin{tabular}{|c c c c c c c c c c|} 
	 \hline
	 C1 & C2 & C3 & C4 & C5 & C6 & C7 & C8 & C9 & C10  \\
	 n & $z_{start}$ & $N_{box}$ & $\sqrt[3]{N_{part}}$ &realisations& No. of snapshots & $r_{nl}^{sim}$ & $r_{nl}^{mf}$ & $r_{nl}^{max}$ & CPU hours\\ [0.5ex] 
	 \hline
	 0.0 & 453 & 512 & 512 & 10 & 12 & 1.25-29.2.3&4.66-22.49 & 58.7 & 3,21,020\\ 
	 -0.5 & 270 & 512 & 512 & 10 & 15 & 1.25-49.4&4.66-38.01 & 42.3 & 3,98,608\\
	 -1.0 & 160 & 512 & 512 & 10 & 15 & 1.25-49.4& 3.58-22.49 & 27.4 & 2,30,320\\
	 -1.5 & 96 & 1024 & 1024 & 1 & 15 & 1.25-49.4&2.76-22.49 & 27.8 & 1,18,800\\ 
	 -1.8 & 70 & 1024 & 1024 & 1 & 15 & 1.25-49.4&1.63-13.31 & 14.5 & 88,998\\
	 -2.0 & 57 & 1536 & 1536 & 1 & 16 & 0.2-10.24&1.25-7.88 & 11.4 & 1,30,743\\
	 -2.2 & 46 & 1536 & 1536 & 1 & 16 & 0.2-10.24& 0.96-2.78& 4.4 & 1,53,520\\
	 \hline

	 \end{tabular}
	 \caption{\label{tab:sim}{\bf Simulation Setup:} C1:Power law power spectrum index for the model, C2:Initial redshift used to start the simulation, C3:Side length of the cubical simulation box, C4:Cube root of the total number of particles put in the simulation, C5:Number of realizations run for the model, C6:Number of snapshots written during the simulation run, C7:Range of the scale of non-linearity($r_{nl}$) covered in the simulation, C8:Range of $r_{nl}$ used to compute the mass function, C9:Maximum limit on $r_{nl}$ considering the finite box size effect, C10:Number of CPU hours taken by the simulation.}
	\end{center}
	\end{table*}
	
We run a suite of dark-matter-only simulations for seven power-law power spectrum initial conditions.
Power-law indices ($n$) of the power spectrum ($P(k)=Ak^n$) are shown in column 1 of table \ref{tab:sim}.
We make use of \textsc{gadget4} \citep{springel2020simulating} to run the simulation suite.
\textsc{gadget4} is run in TreePM configuration with initial conditions generated using second order Lagrangian perturbation theory.
We set the split radius to 2 in grid units. The split radius divides the interactions between particles into two groups, short-range and long-range interactions. The softening length is set to $\epsilon$ = 0.05 grid units for all the runs. We use adaptive time steps with a maximum step size of 0.005. Our chosen parameter values ensure that the force calculation errors are well below 1\%.

EdS is chosen as the background cosmology, where the growing mode growth factor is the same as the scale factor. 
It is the simplest model as threshold collapse overdensity is constant. 
The only scale involved in these models is the scale of non-linearity ($r_{nl}$) introduced by gravity.
It is deﬁned as the scale for which the linearly extrapolated value of the mass variance at a given epoch $\sigma_L(z,r_{nl})$ is unity.
We can therefore identify any epoch in terms of $r_{nl}$ ($r_{nl} \propto a^{2/(n+3)}$).

The average inter-particle separation at the beginning of the simulation is set to unit grid size, i.e., the number of particles in the simulations $N_{part}$ equal to the volume of the simulation in grid units.
Column 3 of table \ref{tab:sim} lists the box size ($N_{box}$) of the simulation, which is also equal to the cube root of the number of particles ($N_{part}$) in the simulation as shown in column 4.
All power spectra are normalized such that $\sigma(a = 1, r_{nl}=8)=1$.
Initial redshift $z_{start}$ for simulation is estimated using epoch at which mass variance at unit length is of the order of $10^{-2}$ shown in column 2 of table \ref{tab:sim}.
It establishes a consistent starting point regarding the fluctuation amplitude of all seven models being compared here. However, it is essential to note that as per \cite{Michaux2020}, systematic errors can arise in simulations due to initial conditions. Therefore, we have verified that the results presented here are not affected by the chosen $z_{start}$.
Column 5 lists the number of realizations run for a given model. 
For the first three models, we use multiple realizations of smaller boxes to account for sample variance.
For others, we use a single realization of larger boxes to cover a broad mass range.

\begin{figure}
    \includegraphics[width=0.45\textwidth]{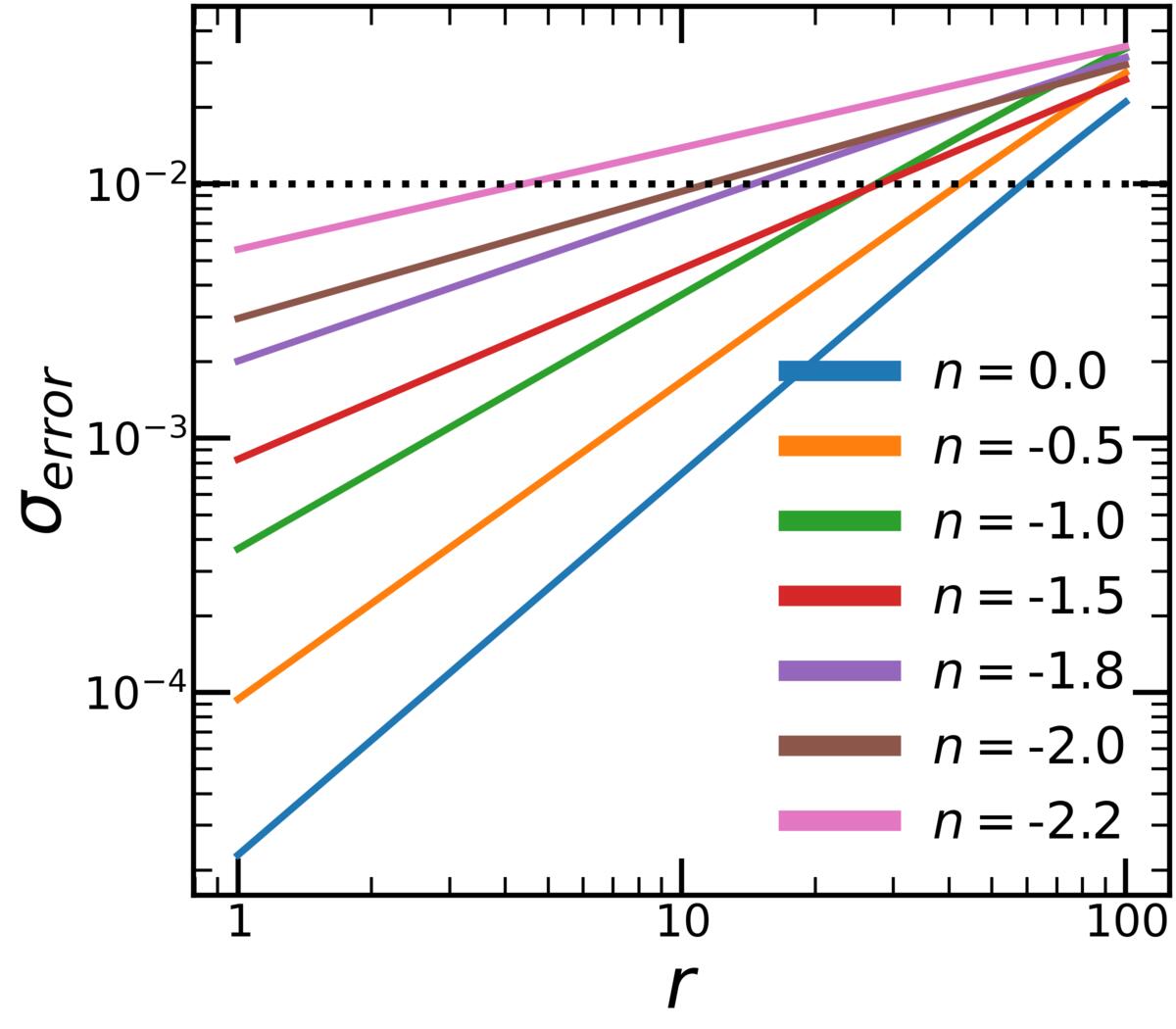}
    \caption{{\bf Finite Box Size Effect:} The plot shows the fractional error in the mass variance due to the finite box size used in our models as a function of length scale in the grid units.
    Different colored solid lines represent the models listed in the table \ref{tab:sim} with power-law power spectrum index n as displayed in the plot legend.
    The dotted black line represents the 1\% mark.}
    \label{fig:sigma_err}
\end{figure}

Finite volume in N-body simulations causes significant errors, as modes greater than the size of the box are not considered while generating initial conditions and during the evolution.
The errors in the mass variance can become arbitrarily large as the power spectrum slope approaches -3.0.
Furthermore, as demonstrated by \cite{Klypin2019}, the finite box size can lead to the propagation of errors in the mass function for LCDM cosmologies. These factors should be considered when determining the appropriate box size and the regime in which the results can be considered reliable.
We use the prescription of \citet{bagla2006effects} for deciding the box size and the range of $r_{nl}$.
We set the tolerance limit at $ 1 \%$ error in mass variance at the scale of non-linearity.
Figure \ref{fig:sigma_err} shows fractional error in the mass variance of the initial density field as a function of length scale in grid units, calculated using the above prescription.
Various colored solid lines represent listed models in table \ref{tab:sim}, whose power law indices are as displayed in the figure.
As we move towards more steep slope models, larger box sizes need to be selected to keep errors below 1\%.
We have confirmed the stability of our results against variations in the box size.
Column 9 of table \ref{tab:sim} lists the maximum limit on the scale of non-linearity within our criteria.
Column 7 lists the entire $r_{nl}$ range covered to take outputs in simulations.
Epochs are selected such that increments in $\ln r_{nl}$ are a constant.
We set the spacing to the minimum value required to consider outputs independent of each other. 
Column 8 lists $r_{nl}$ range used to calculate the mass function.

Column 10 lists the CPU time in hours taken by simulations to run.
In total, the simulations have taken 1.5 million CPU core hours.

\subsection{Self-similar evolution}
\label{subsec:self-similar}

\begin{figure*}
    \includegraphics[width=1.\textwidth]{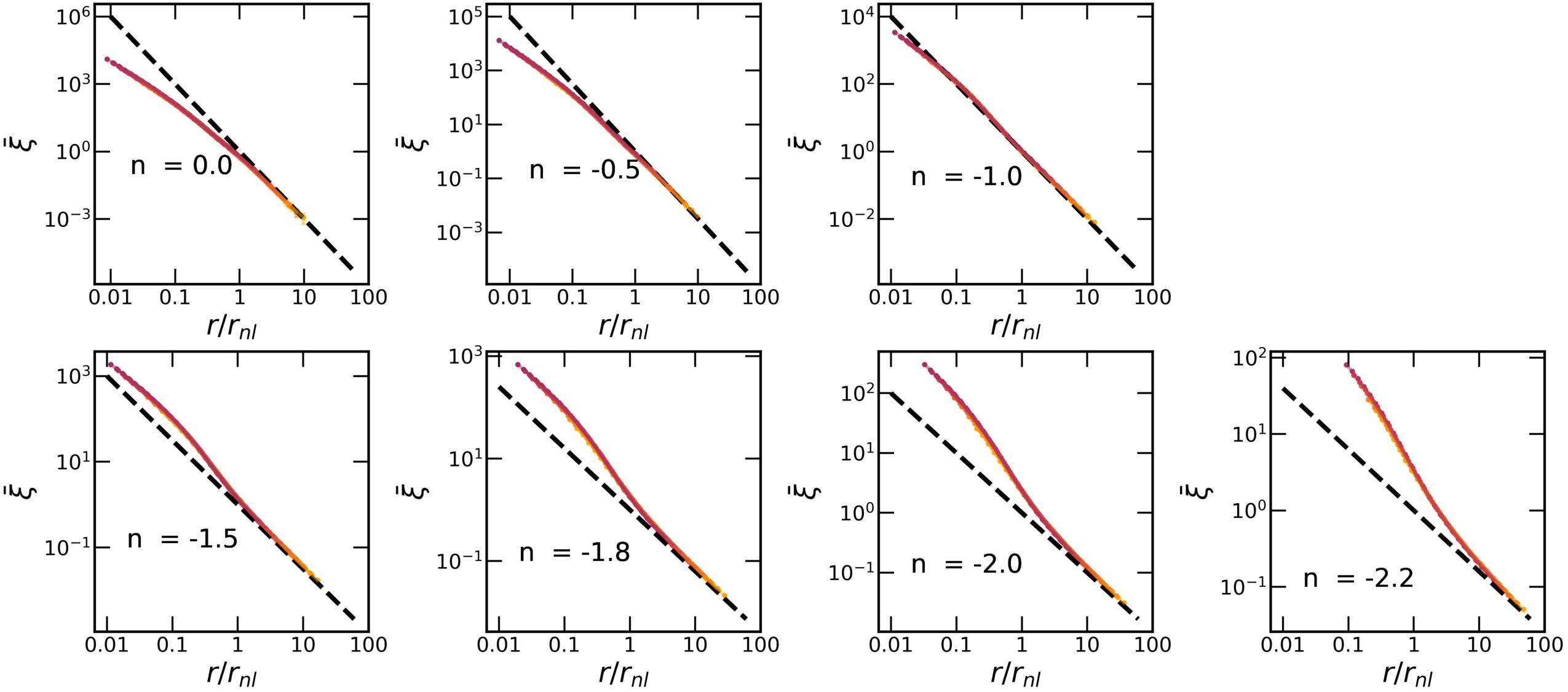}
    \caption{{\bf Self-Similarity:} Figure shows the volume averaged two point correlation function $\bar{\xi}$ scaled with non-linearity scale ($(r_{nl}$), for the all models listed in table \ref{tab:sim}.
    Power law power spectrum index n is displayed on each panel to identify the models.
    Different colored solid lines represent $\bar{\xi}$ at different epochs used to calculate the mass function.
    The dashed black line represents the linear theory prediction for $\bar{\xi}$.}
    \label{fig:corr_1}
\end{figure*}

Finite box size effects induce error in non-linear spectra, where Fourier mode coupling becomes dominant.
During this process, there is a transfer of power from large to small scales; the effect becomes significant as the spectral index approaches -3, as there is equal power in all modes in this limit.
We are able to avoid errors arising from such effects by working with self similar (power law) models and EdS background.  
This approach has been used in many simulation studies to ensure that results are free from errors arising due to finite box size or other effects, e.g., \citet{Maleubre:2022swk}.
We require self similar evolution of clustering across epochs used for the analysis in this paper. 
We check for self similarity by using the volume-averaged two point correlation function $\bar{\xi}$.
We draw random samples of particles from the simulation output to compute $\bar\xi$. 
The range of $r_{nl}$ used in this work for each model is given in column 8 of table \ref{tab:sim}.
Python module \textsc{corrfunc} (\citet{sinha2020corrfunc}) is used to calculate the correlation function.
We show the results in figure \ref{fig:corr_1}.
Different colored solid lines represent $\bar{\xi}$ scaled with $r_{nl}$.
The dashed black line represents linear theory prediction for the same quantity, as in equation \ref{eqn:xi}.
The self-similar character of these models is evident as $\bar{\xi}$ plotted for various epochs overlaid to form the single curve as a function of $r/r_{nl}$.
There are three main regimes that can be seen in the $\bar{\xi}$ curves: linear($r/r_{nl}\gg 1$), quasi-linear($r/r_{nl}\approx1$) and non-linear($r/r_{nl}\ll 1$).`
During the evolution of the matter density field, small scale collapses first enter into quasi-linear and then non-linear regime.
Growth in the quasi-linear regime remains close to the linear rate for $n=-1$, and it is faster for $n < -1$ and slower for $n>-1$. Thus the slope of $\bar\xi$ for all models ultimately approaches $-2$ in a non-linear regime, the same as that for the $n=-1$ model \citep{1997MNRAS.286.1023B}. 
It can be seen around the $n=-2$ models.

\begin{equation}
    \bar{\xi}(r) \approx \left( \frac{r}{r_{nl}}\right) ^{ -(n+3)}
    \label{eqn:xi}
\end{equation}

\subsection{Finding halos}
\label{subsec:halo_finding}

The inbuilt \textsc{fof-subfind} \citep{springel2001populating} algorithm of \textsc{gadget4} is used to construct the halo catalogue with a linking length equal to 0.2 and all its default parameters.
It first identifies Friends-Of-Friends (FOF) \citep{Davis1985} groups in given distribution and then decomposes each found object into substructures using the excursion set algorithm. 
We only consider isolated halos to compute the halo mass function.
The halo-finding approach(SO/FOF) uses spherical overdensity to define halo masses, and hence SO masses differ from the FOF masses. 
The finite box size effect and discreteness noise can affect halo masses at the high and low mass end, respectively. 
Care is required to ensure that such effects do not creep into our analysis.  
We describe our approach in the following section.

\section{Results}
\label{sec:results}

Here we present the halo mass function from our simulations of power-law models and compare them with theory predictions.
First, we bin the available halo catalog with mass to compute the halo mass function.
We construct adaptive bins in mass such that each bin contains a fixed number of halos.
This gives us a halo count per mass bin, $dn/d\ln m$.
With this, equation \ref{eqn:dn} can now be rewritten as:
\begin{equation}
    f(\nu) = \frac{6}{n+3} \frac{m}{\bar{\rho}} \frac{dn}{d\ln m}
    \label{eqn:fnu}
\end{equation}
Note that $\bar{\rho} =1$ in units used here.
Using the following relation, we arrive at Equation \ref{eqn:fnu}.
\begin{equation}
    \sigma(m) =\left( \frac{m}{m_{nl}}\right)^{ - \frac{n+3}{6}}
    \label{eqn:sigma}
\end{equation}
where, $m = 4 \pi r^3 /3$.

We fit ST mass function to our data using $\chi^2$ minimization with p and q as free parameters.
To determine the errors in our analysis, we considered the Poisson errors associated with the number of halos in each bin. Furthermore, we assessed the suitability of our fits by employing Jackknife and Sample Variance errors. We verify that using Poisson errors is a conservative approach and that the outcomes obtained using this method are equally robust as those obtained using other error types.
In subsequent sections, we discuss mass function plots, tolerance to the filtering criteria, the 
dependence of ST parameters on power law n, and epoch dependence.

\subsection{Mass function}
\label{sec:mf_result}

\begin{figure*}
    \includegraphics[width=0.99\textwidth]{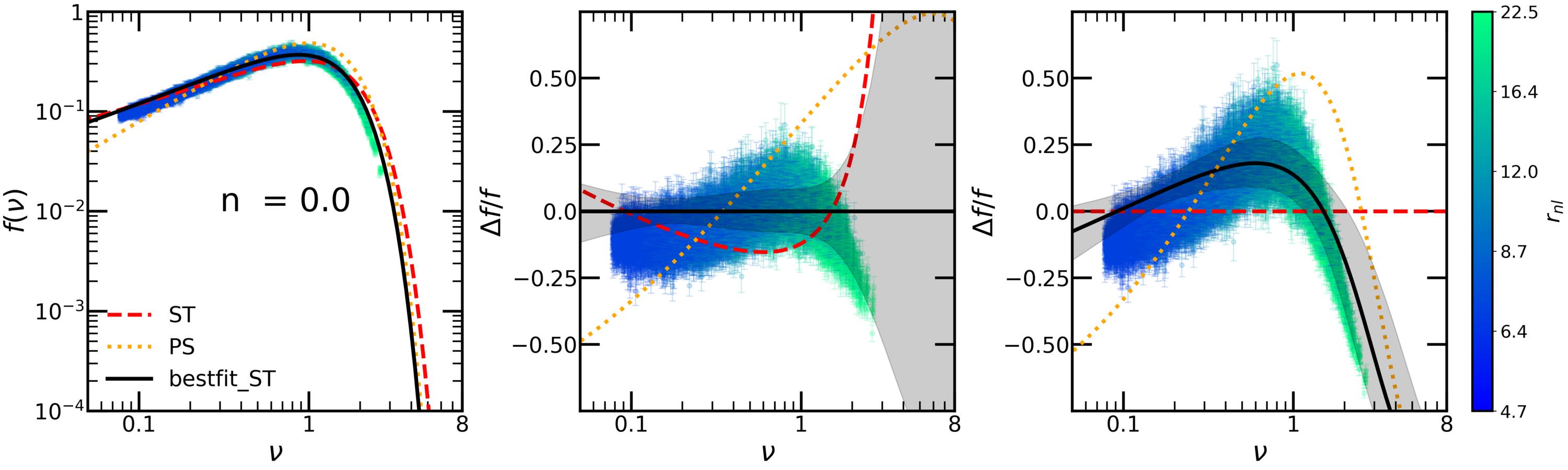}
    \includegraphics[width=0.99\textwidth]{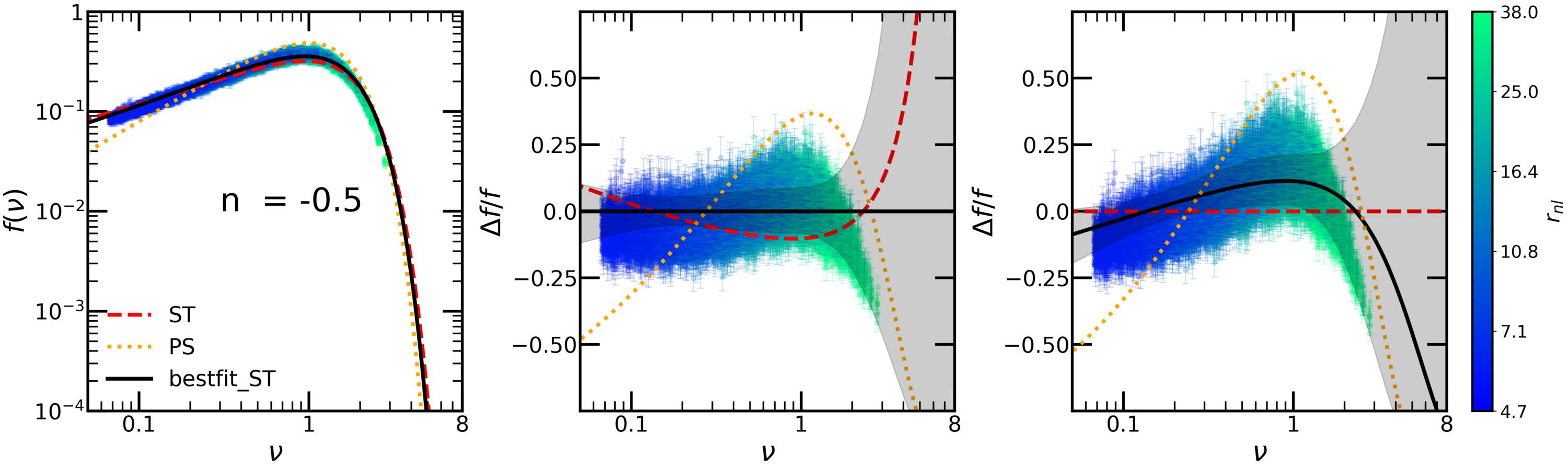}
    \includegraphics[width=0.99\textwidth]{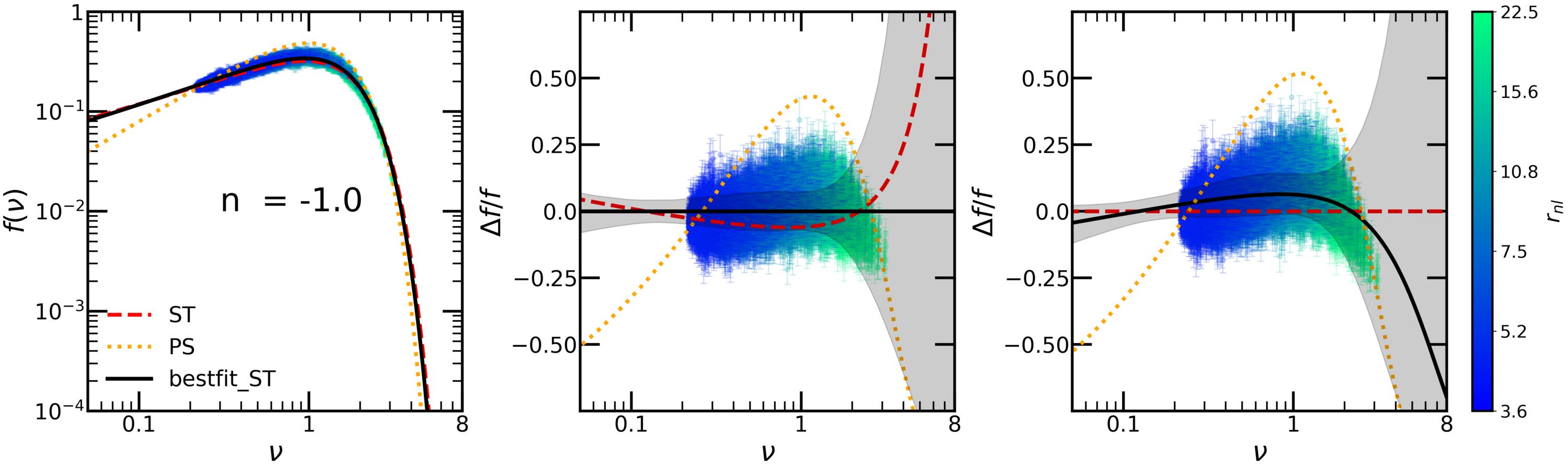}
    \includegraphics[width=0.99\textwidth]{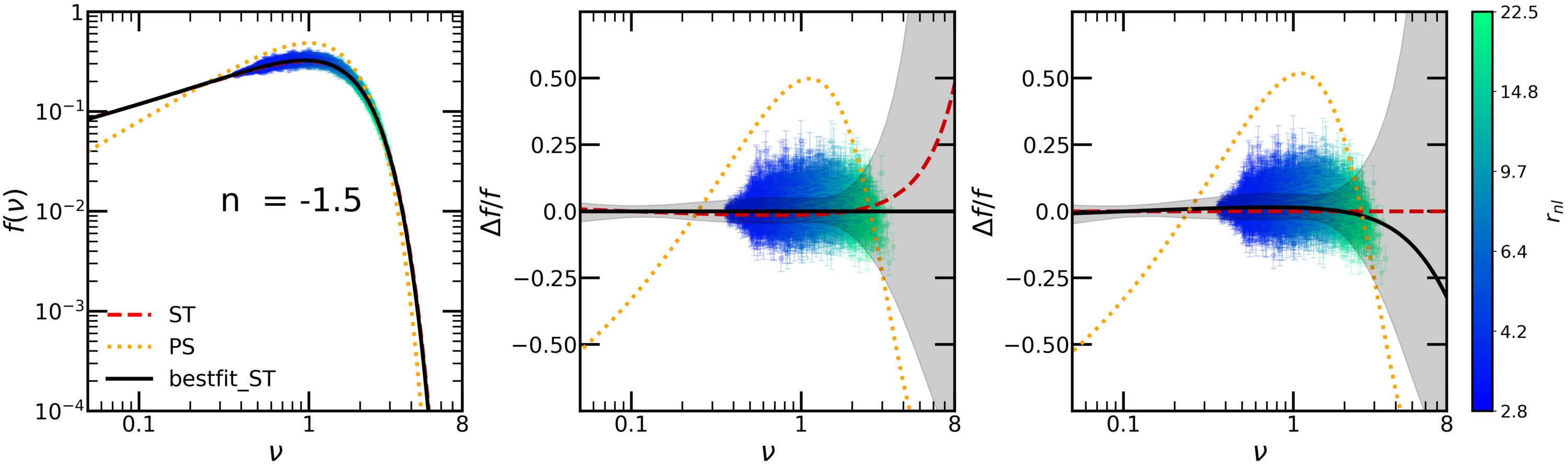}
    \caption{\textbf{Mass function:} The figure shows the mass function from the simulations (left column), their residuals after fitting the Sheth-Tormen function (middle column) and residuals with respect to standard Sheth-Tormen function (right column). Rows 1,2,3 and 4 are for power law indices 0.0, -0.5, -1.0 and -1.5 respectively. Dotted orange and dashed red lines represent the Press-Schechter and the Sheth-Tormen mass function, respectively. Solid black lines represent the best fit Sheth-Tormen mass function after $\chi^2$ analysis. Blue-green data points show the mass function calculated from the simulations. Data points color coded with respect to $r_{nl}$, color changes from green to blue as $r_{nl}$ increases as shown in adjacent color-bar. Gray filled area in the middle and right panel shows the 1$\sigma$ confidence interval for the fit.
    }
    \label{fig:mf_1}
\end{figure*}

\begin{figure*}
    \includegraphics[width=0.999\textwidth]{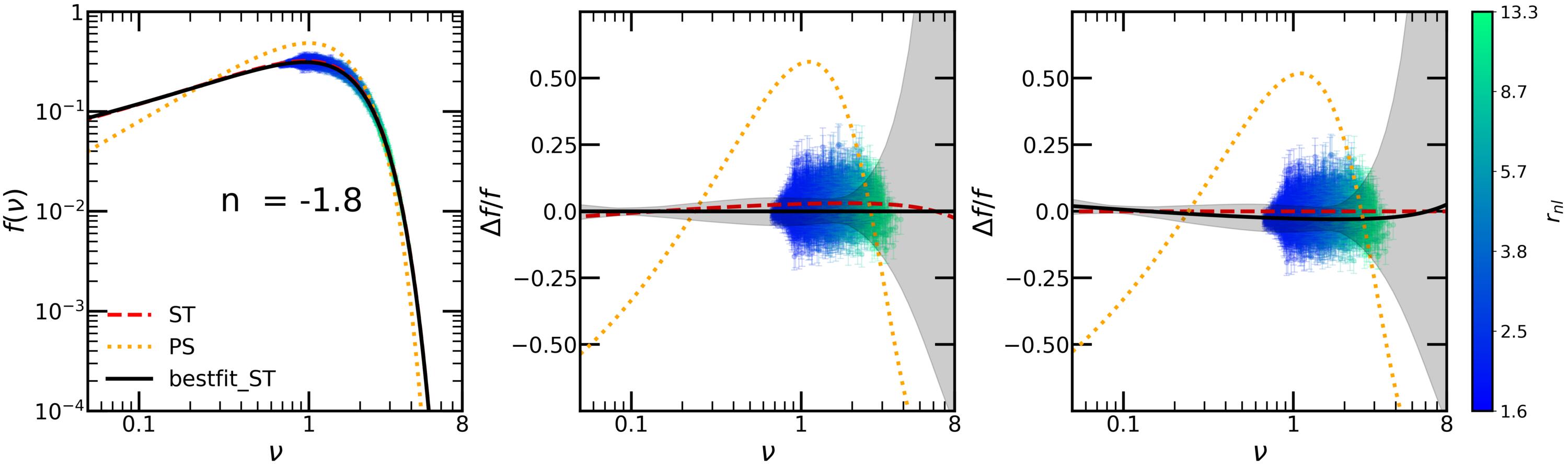}
    \includegraphics[width=0.99\textwidth]{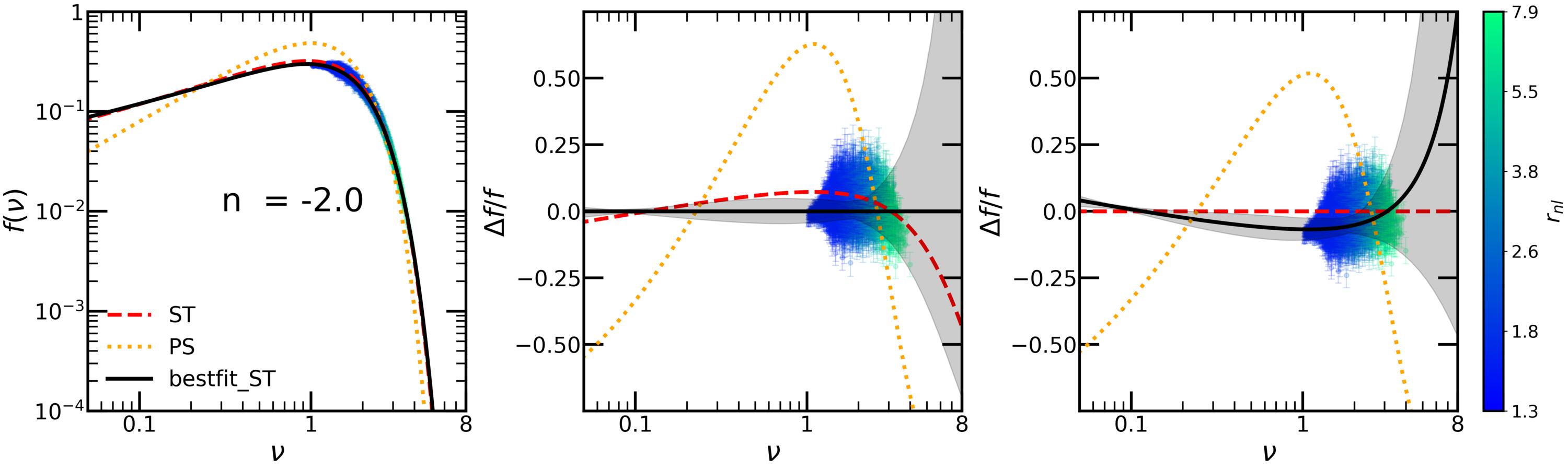}
    \includegraphics[width=0.99\textwidth]{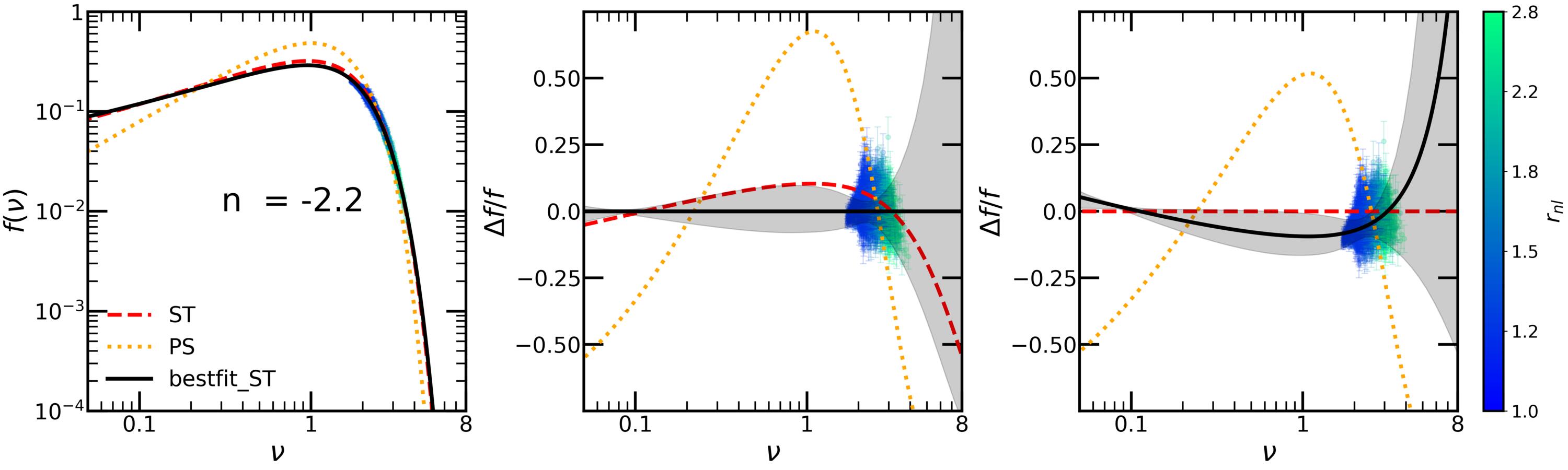}
    \caption{\textbf{Mass function:} The figure shows the mass function from the simulations (left column), their residuals after fitting the Sheth-Tormen function (middle column) and residuals with respect to standard Sheth-Tormen function (right column). Rows 1,2,3 and 4 are for power law indices -1.8, -2.0 and -2.2 respectively. Dotted orange and dashed red lines represent the Press-Schechter and the Sheth-Tormen mass function, respectively. Solid black lines represent the best fit Sheth-Tormen mass function after $\chi^2$ analysis. Blue-green data points show the mass function calculated from the simulations. Data points color coded with respect to $r_{nl}$, color changes from green to blue as $r_{nl}$ increasesas shown in adjacent color-bar. Gray filled area in the middle and right panel shows the 1$\sigma$ confidence interval for the fit.}
    \label{fig:mf_2}
\end{figure*}

We define SO halo mass $M_{\Delta c}$ as the mass within a radius $r_{\Delta}$ inside which the average density is $\Delta$ times the critical density.
We use masses with $\Delta=200$ to compute the mass function.
$M_{vir}$ is the mass corresponding to the virial overdensity($\Delta=18\pi^2$).
We only consider halos with more than $100$ particles to avoid discreteness noise.
The fixed count for bins is chosen to be $300$ halos per bin.
At the high mass end, the number counts of halos drop sharply and the fixed bin size requirement can create problems. 
The halo count at the high mass end is also affected strongly by cosmic variance. 
To remedy this, we remove the most massive $0.1\%$ halos during the computation of the mass function.
We discuss our choice of these values in the following subsection, along with variation in the best fit values as we vary these thresholds/parameters.

Figures \ref{fig:mf_1} and \ref{fig:mf_2} show mass function results.
The left panel shows mass function, while the middle and the right panel show residual of the data with respect to best fit ST and standard ST, respectively.
Dotted orange and dashed red lines represent PS and ST mass functions, respectively.
A solid black line represents the best fit ST mass function after $\chi^2$ analysis.
Blue-green data points show mass function calculated from simulations.
Data points are color coded for $r_{nl}$, and color changes from green to blue as $r_{nl}$ increases, as shown the in adjacent color-bar.
The gray filled area in the middle and right panel shows 1$\sigma$ confidence interval around the best fit.
We find the following points noteworthy in these plots:
\begin{itemize}
  \item 
  The best fit curve (black) in the left panels or deviation of residual PS and ST curves in the middle panels show a systematic pattern in the shape of mass function and its variation with power-law power spectrum index $n$ can be seen.
  \item
  The right panels show data points, and the data point fits moving away from standard ST systematically. This points inadequacy of ST to explain the theory of mass function fully.
  \item
  We use adaptive binning(fixed number of halos per bin) to construct mass function. We verify that it gives better justification to the high mass end as fits are less biased to the low mass end here compared to fixed bin width in $\log_{10} (\nu)$. However, it is still insufficient, as data points show deviations at the high mass end, especially for n = 0, -0.5, -1.0. Probably one needs additional theory parameters to explain the high mass end.
  \item 
  The range of the mass function is smaller as we move towards models with more negative power law power spectrum indices. We have not evolved simulations with these indices over a broad range of epochs considering finite box size effects. A smaller range in $\nu$ for $n = 0.0$ compared to $n = -0.5$ is due to a smaller range of $r_{nl}$. Evolution in the non-linear regime becomes dramatically slow as $n$ approaches zero.
  \item 
  We notice n=-1.5 as an index around which deviation of the standard ST function from the best fit shifts sign. This may be related to the effective index in the simulations where the ST function was calibrated.
  \item 
  As we progress towards late time, characteristic mass $M_*$ increases, leading to a shift of $\nu$ interval towards low values(data points changing color from green to blue).
  \item 
  $\chi^2$ is biased by the low mass end as the data points are clustered there; we see a deviation of data points at the high mass end from the best fit curve, especially for n = 0.0,-0.5 and -1.0.
  \item 
  Choosing a fixed number of halo counts per mass bin leads to equal Poisson error bars; however, we see some data points have smaller error bars at the low mass end. This is because there are more halos in a particular mass interval than the fixed count value.
  \item
  Low mass end behaviour of ST mass function is decided by p, while high mass end exponential cutoff depends on q. This leads to a large variation in mass function at the low(high) mass end due to q(p) change. In the transition region, the range of p and q in the fits is so that the ST mass function does not show much variation. Due to this, we see tapering of 1$\sigma$ band around $\nu \approx 0.2$.
  
\end{itemize}

\subsection{Dependence of Sheth-Tormen parameters of power law index and epoch}
\label{sec:n_rnl_dep}

\begin{figure*}
    \centering
        \begin{subfigure}[b]{0.6\textwidth}
        \includegraphics[width=.99\textwidth]{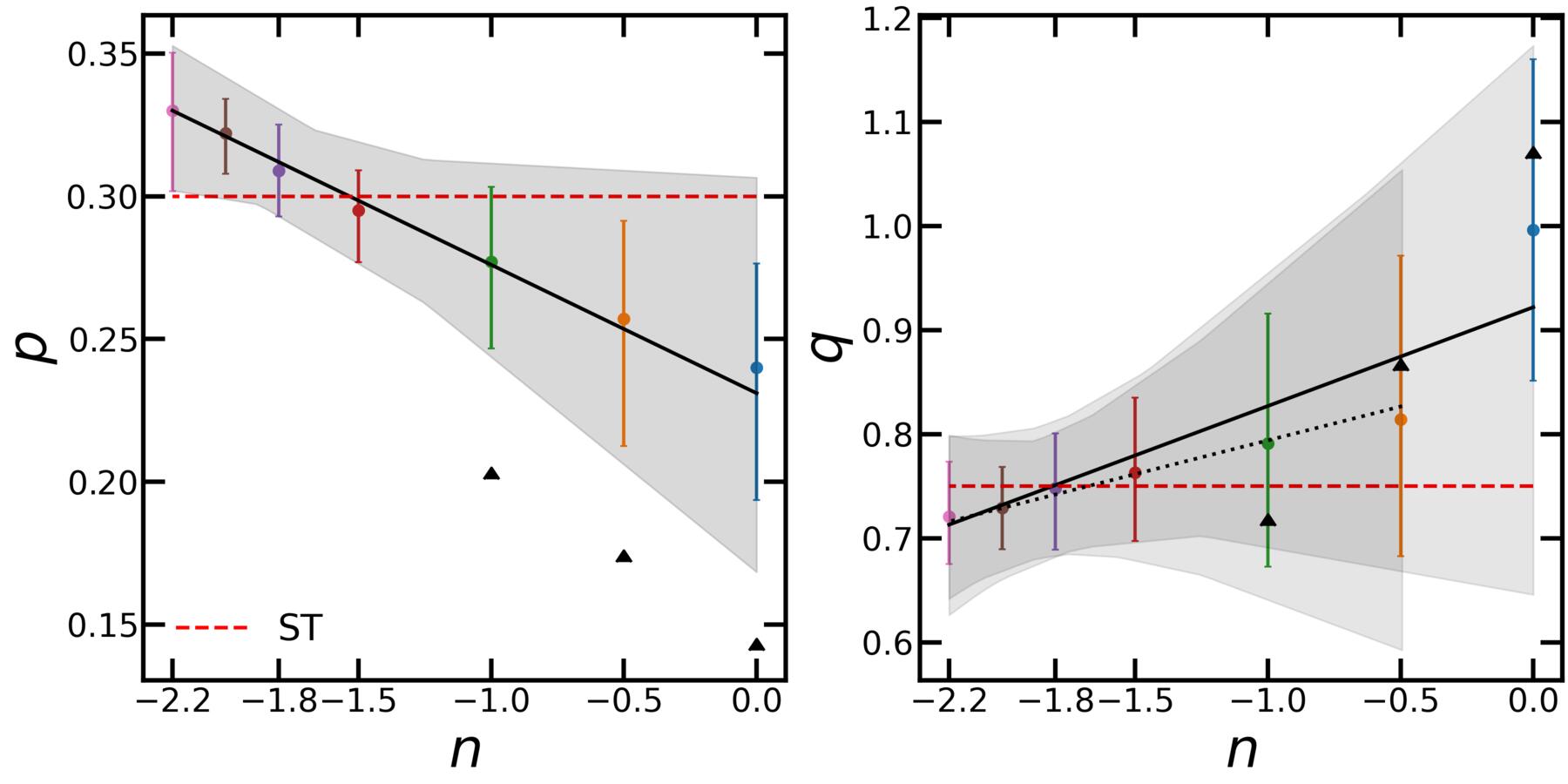}
        \end{subfigure}
        \begin{subfigure}[b]{0.3\textwidth}
        \includegraphics[width=.99\textwidth]{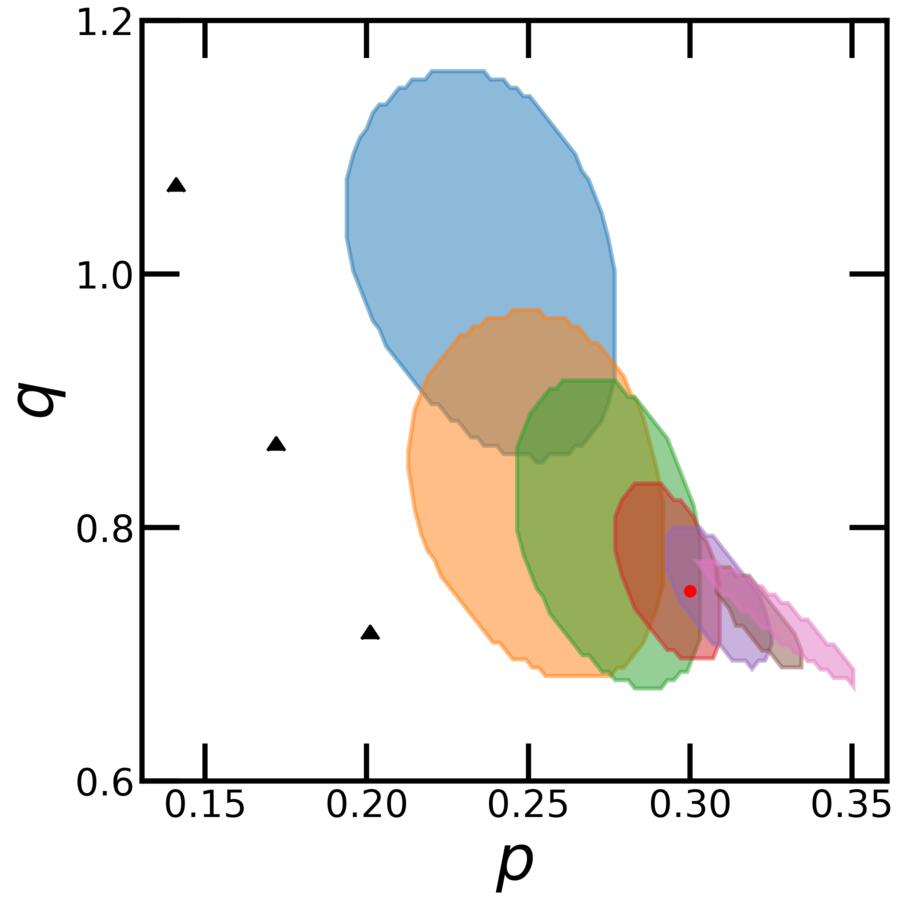}
        \end{subfigure}
    \caption{{\bf Power spectrum dependence of the Sheth-Tormen parameters:} Figure shows the variation of the Sheth-Tormen parameters with the power law index (p:left panel, q:right panel). Errors in the p,q corresponds to the 1$\sigma$ confidence in the fitting. A solid black line represents a linear fit to the scatter with the gray area as 1$\sigma$ confidence of this fitting. The right panel shows the 1$\sigma$ contours of the mass function fittings in the pq-space. Dashed red lines in the left and middle panels show standard ST parameters, the corresponding red marker is shown in the right panel. The dotted black line in the middle panel shows a revised linear fit to the scatter(see text for details). Triangular data points show best fit values from \protect\cite{bagla2009mass}. }
    \label{fig:pq_n}
\end{figure*}

We looked at the variation of best fit ST parameters p and q with power spectrum index n.
We find the following approximate n dependence of p and q after fitting straight lines.
The choice of a linear relation appears to be a good first approximation as seen in figure \ref{fig:pq_n}.

\begin{equation}
  \begin{array}{l}
  p(n) \simeq -0.045 n+0.231\\ 
  q(n) \simeq 0.095 n+0.922
  \end{array}
    \label{eqn:pq_n}
\end{equation}

The left and the middle panel of figure \ref{fig:pq_n} show the relation between ST parameters with power-law power spectrum index $n$.
Errors in $p$ and $q$ are derived from $1\sigma$ confidence in the fitting.
Solid black lines show linear fit (equation \ref{eqn:pq_n}) to $p$ and $q$ scatter.
Gray filled areas show $1\sigma$ confidence for the linear fits.
Dashed red lines in left and middle panels show standard ST parameters.
The right panel of figure \ref{fig:pq_n} show $1\sigma$ contours of mass function in pq-space.
Triangular data points show best fit values from \cite{bagla2009mass}.
This shows a correlation between ST parameters.
The overall plot shows a clear trend of ST parameters with power spectrum index $n$.
Thus we have clear evidence in these simulations of the dependence of the parameters of the mass function on the slope of the initial power spectrum. 
This indicates a departure from the universality of mass functions. 
This is a significant finding and requires theoretical investigations.  
Understanding this will lead to improved theoretical modelling of the collapse of halos and the theory of mass functions.

We see n=0 as an outlier with the linear fit.
If we eliminate this data point, we get revised linear fits as in equation \ref{eqn:pq_n_revised}.
Eliminating the n=0 data point does not affect the p trend at all but affects the slope of the q line significantly.
Revised fit and corresponding 1$\sigma$ confidence interval is shown in the middle panel(black dotted line) of figure \ref{fig:pq_n}.
It is a better predictor for LCDM models, as we will see in section \ref{sec:lcdm}.

\begin{equation}
  \begin{array}{l}
  p(n) \simeq -0.045 n+0.231\\ 
  q(n) \simeq 0.065 n+0.859
  \end{array}
    \label{eqn:pq_n_revised}
\end{equation}

\begin{figure}
    \includegraphics[width=.47\textwidth]{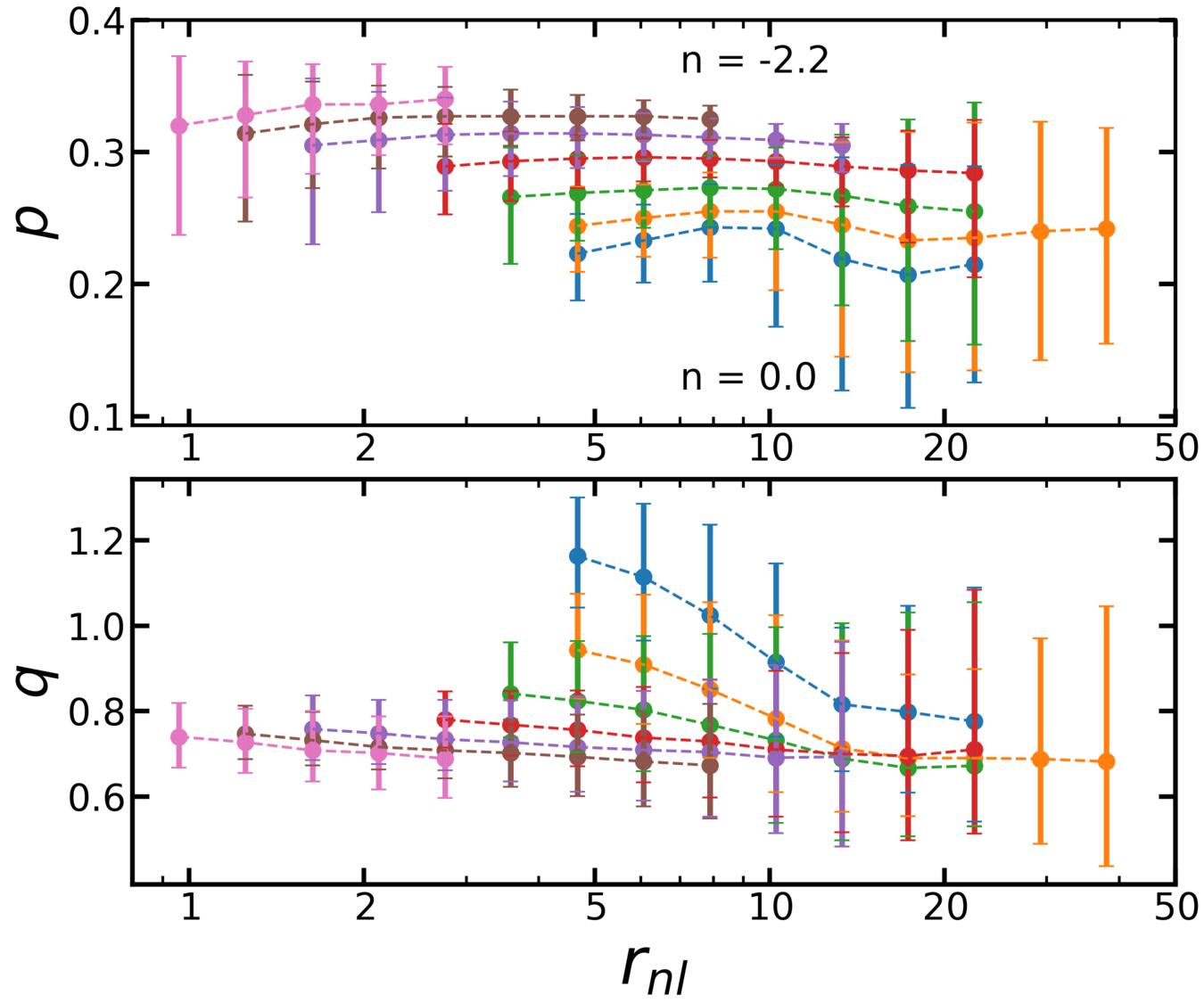}
    \caption{{\bf Epoch dependence of the Sheth-Tormen parameters:} The figure shows the variation of the Sheth-Tormen parameters (p:upper panel, q:lower panel) with the scale on non-linearity ($r_{nl}$). Errors in the p,q corresponds to the 1$\sigma$ confidence in the fitting. Different colors used for the scatters represent different models with the power law n as shown in the legend of figure \ref{fig:sigma_err}.
    }
    \label{fig:pq_rnl}
\end{figure}

We fit epochs available for a model separately to investigate the time dependence of the shape of the mass function.
Figure \ref{fig:pq_rnl} shows variation of p and q with $r_{nl}$.
$p$ tend to remain stable with $r_{nl}$.
$q$ varies with power law index  $n$; variation in $q$ with $r_{nl}$ tends to increase as $n$ approaches zero.
This perhaps indicates the presence of some transient effects.
Thus we conclude that the shape of the mass function does not vary with time in a significant manner for n<-1.

Trends in p and q presented by \citet{bagla2009mass} match qualitatively with our findings.  
In the present study, we have explored more models and used larger simulations.  
We have also studied the tolerance towards various parameters used in the study as discussed in the following section. 

\subsection{Tolerance to parameters}
\label{sec:mf_tolerance}

\begin{figure*}
    \includegraphics[width=.9\textwidth]{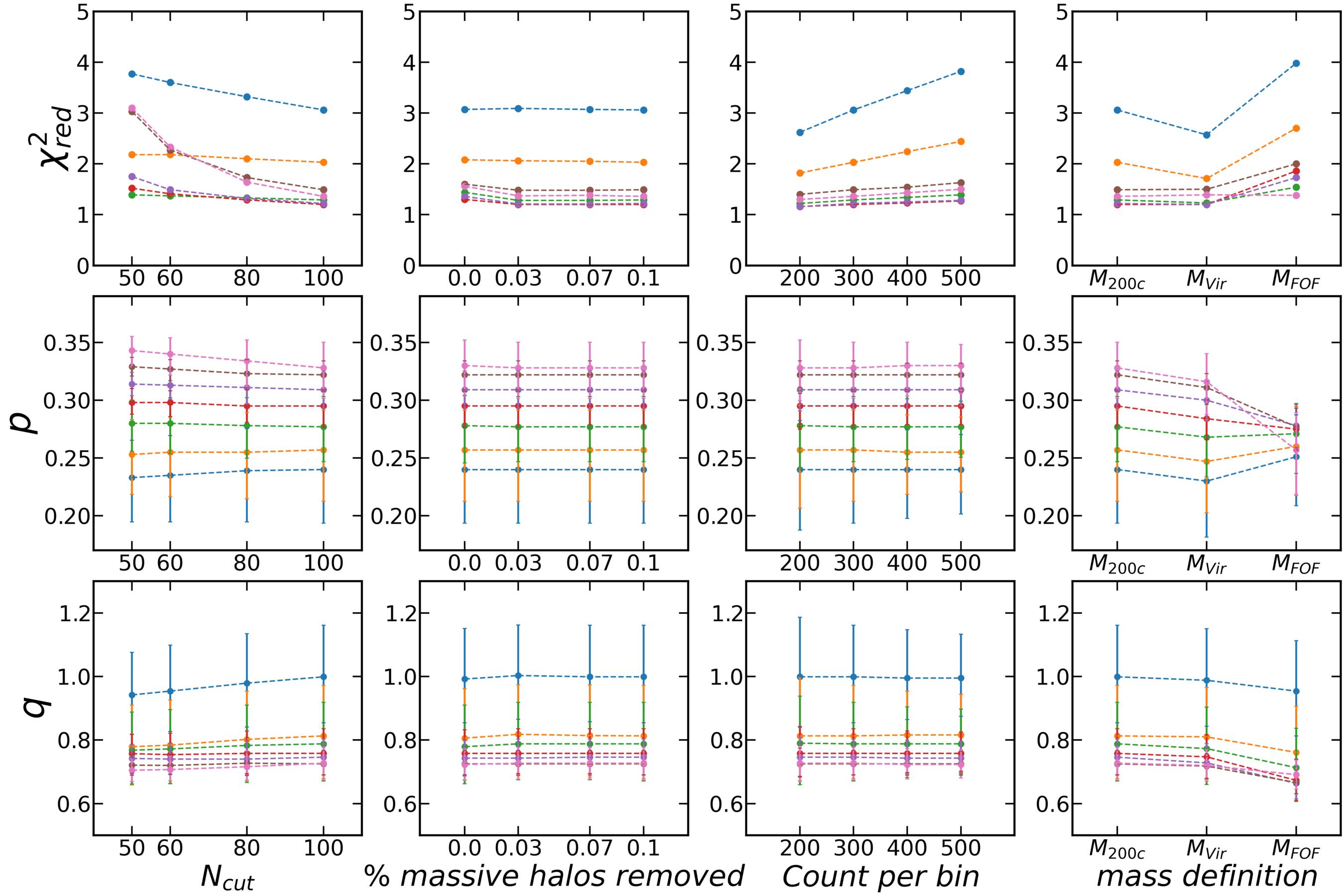}
    \caption{\textbf{Tolerance to parameters:} Effect of parameters like $N_{cut}$ (column 1), percentage of massive halos removed from halo catalog (column 2), number of halos in each bin (column 3), and mass definitions (column 4) on  $\chi^2$ (row 1) and fitting parameters (row 2, row3).
    Different curves in each panel represent different power law models, colors used for models are the same as shown in figure \ref{fig:sigma_err}.}
    \label{fig:tol}
\end{figure*}

In this subsection, we present how the criteria used to prepare the halo catalog obtained from simulations to affect the mass function and the best fit.
We test the criteria used: removal of small mass halos, removal of massive halos, number of halos in each bin, and halo mass definitions.

The parameters that we vary are:
\begin{itemize}
    \item 
    $N_{cut}$  is the minimum number of particles a halo must have for us to consider it in the halo catalog used for fitting the mass function.
    \item
    Most massive halos are rare and the numbers are strongly affected by cosmic variance.  Thus we remove a small fraction of halos from the high mass end from the catalog and carry out fitting the mass function in the remaining bins.  This is the second parameter used in the calculations.
    \item
    The number of halos in an individual bin affects the size of the error bars.  This is the third parameter that we have used. 
    \item
    There are multiple ways to define the mass of halos.  We consider three($M_{200c}, M_{vir}, M_{FOF}$) possibilities in this paper.  We study the tolerance of mass function to the choice of mass definition as well.
\end{itemize}
The fitting procedure gives us the best fit values of the mass function parameters $p$ and $q$, and the reduced $\chi^2$.  We study the variation of these three with the choices we make and the results are shown in figure \ref{fig:tol}.
We have plotted curves for each of the power law models in each panel.  
Columns in this figure correspond to our choices, and rows correspond to the computed values of $\chi^2_{red}$, $p$ and $q$. 
We note that in most panels, the values of these three parameters are stable, indicating that our choices are not critical and the results we are presenting do not depend in any significant manner on the choices.
Specific panels that require a discussion are as follows:
\begin{itemize}
    \item 
    There is an increase in $\chi^2_{red}$ with the number of halos per bin for models with a steeper slope of the power spectrum. This appears to be caused by a reduction in the number of bins.  Interestingly, there is no variation in values of other parameters $p$ and $q$.
    
    As the number of counts in a bin becomes very large, statistical fluctuations in the data become small compared to the bin size. This led to a systematic shift in the last bin. Hence we restrict our analysis to 500 halos per bin.
    \item
    The estimated variation of $p$ with the power law index is very small if we use $M_{FOF}$.  Surprisingly, there is no corresponding effect in a variation of $q$ with the index $n$.  Notably, trends in $p$ and $q$ are intact if the $n=-2.2$ model is ignored.

\end{itemize}

\section{Going beyond power law and EdS}
\label{sec:lcdm}

\begin{figure*}
\begin{subfigure}[b]{0.45\textwidth}
 \includegraphics[width=.95\textwidth]{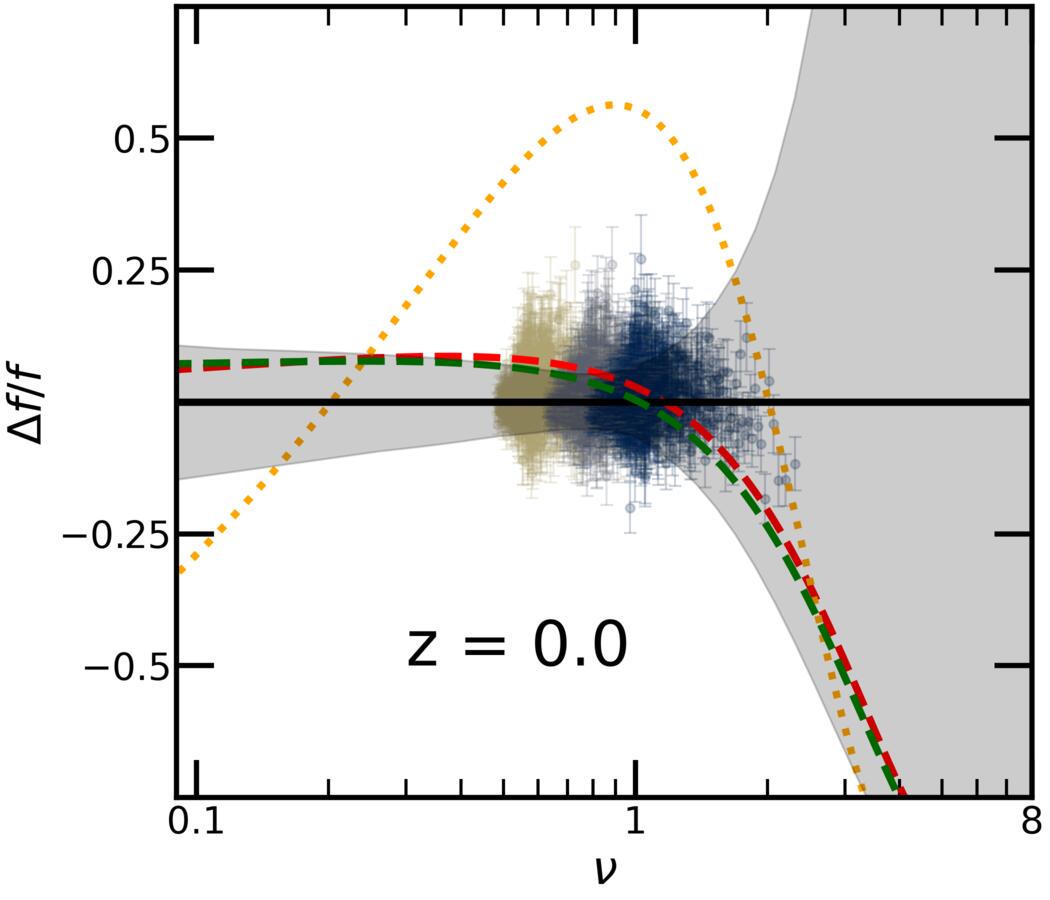}
  \end{subfigure}
      \begin{subfigure}[b]{0.45\textwidth}
 \includegraphics[width=.95\textwidth]{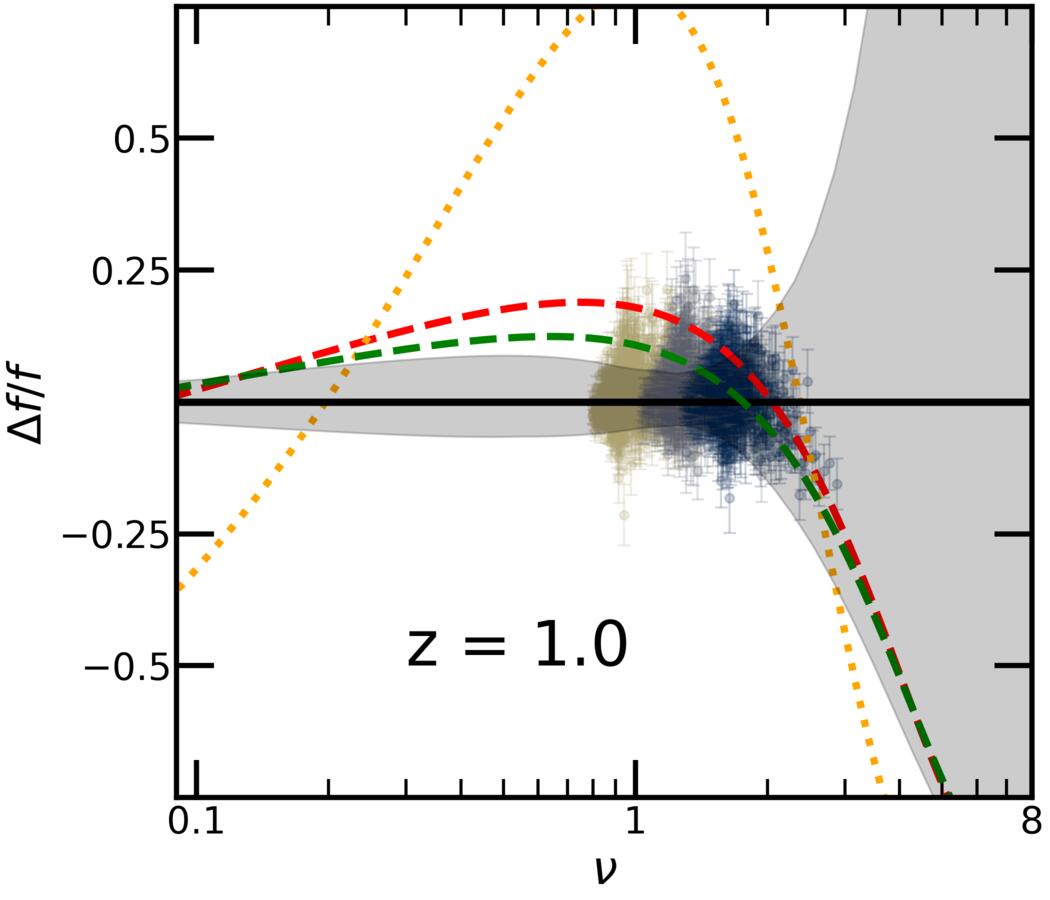}
  \end{subfigure} \\
      \begin{subfigure}[b]{0.45\textwidth}
 \includegraphics[width=.95\textwidth]{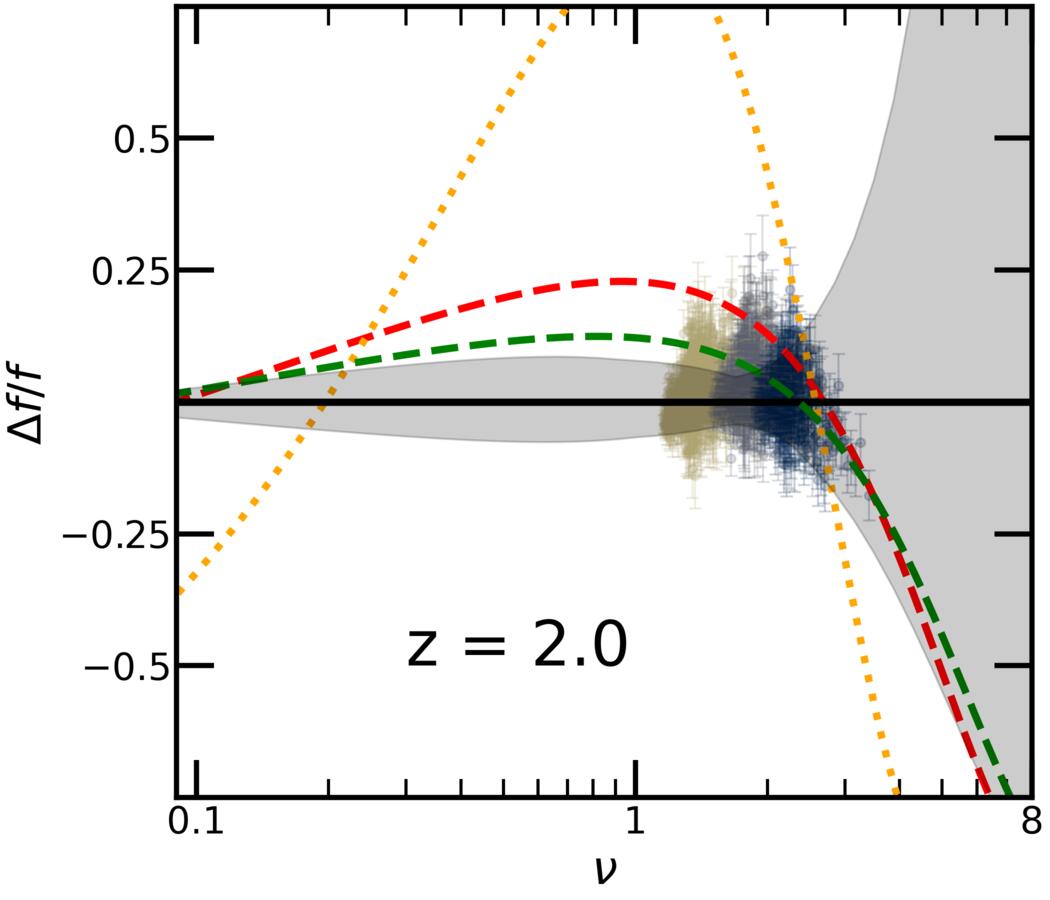}
  \end{subfigure}
      \begin{subfigure}[b]{0.45\textwidth}
 \includegraphics[width=.95\textwidth]{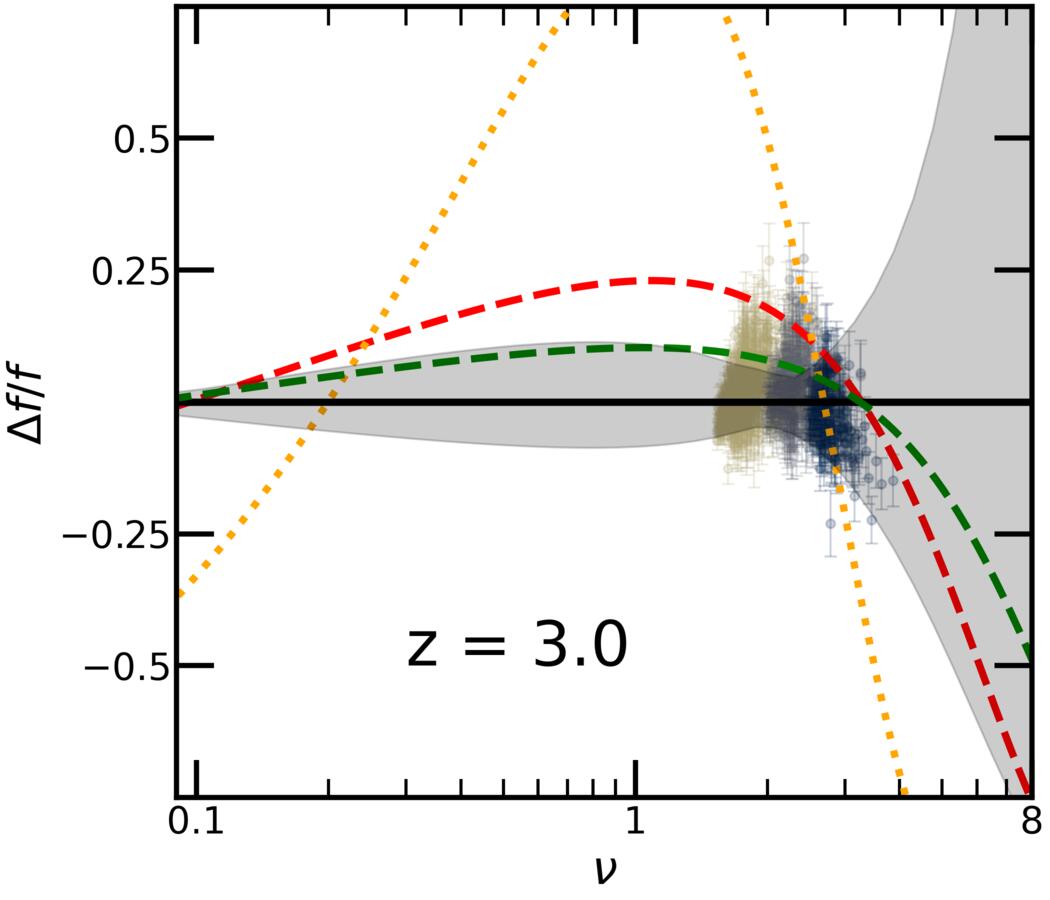}
  \end{subfigure}
  \caption{{\bf Residue of mass function for LCDM models :}  The figure shows the residue of the mass function after fitting the Sheth-Tormen function to LCDM simulation data. Different panels are for different representative epochs. Dotted orange and dashed red lines represent the Press-Schechter and the Sheth-Tormen mass function, respectively. Dashed green lines represent equation \ref{eqn:pq_n_revised}. Solid black lines represent the best fit Sheth-Tormen mass function after $\chi^2$ analysis. Brown-gray data points show the mass function calculated from the simulations. Three different colors of data points represent different simulations used for analysis. Gray filled area shows the 1$\sigma$ confidence interval for the fit.}
    \label{fig:mfl}
\end{figure*}

\begin{figure}
    \includegraphics[width=.47\textwidth]{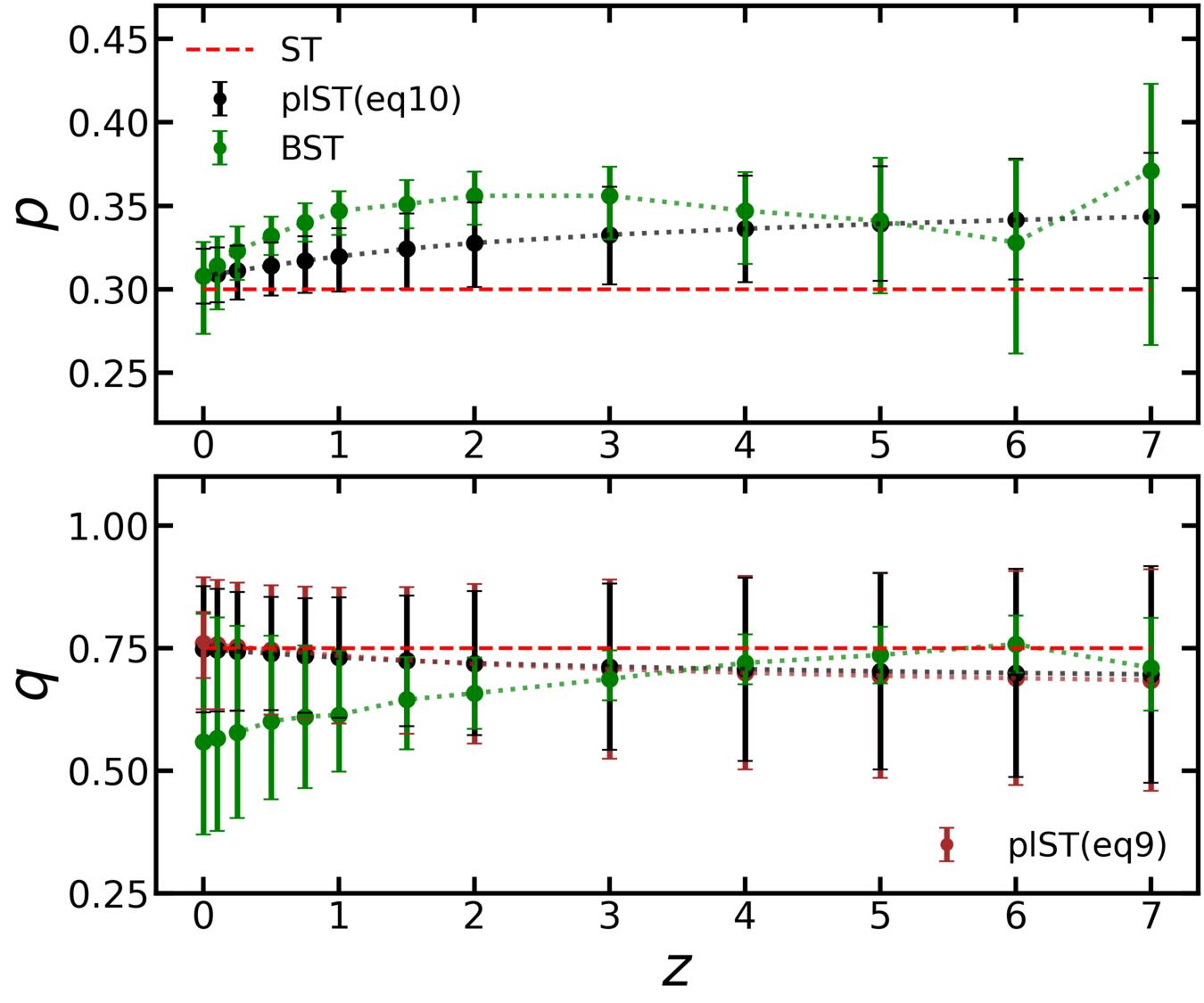}
    \caption{{\bf Sheth-Tormen parameters in LCDM cosmology:} The figure shows the variation of the Sheth-Tormen parameters (p:upper panel, q:lower panel) with the redshift for LCDM model. Green scatter represents best fit ST values while the black and brown data points represent predictions from power law models with respect to equation \ref{eqn:pq_n_revised} and \ref{eqn:pq_n} respectively. Errors in the p,q corresponds to the 1$\sigma$ confidence. Dashed red lines show standard Sheth-Tormen parameters.}
    \label{fig:lcdm_pq}
\end{figure}

\begin{figure}
    \includegraphics[width=.47\textwidth]{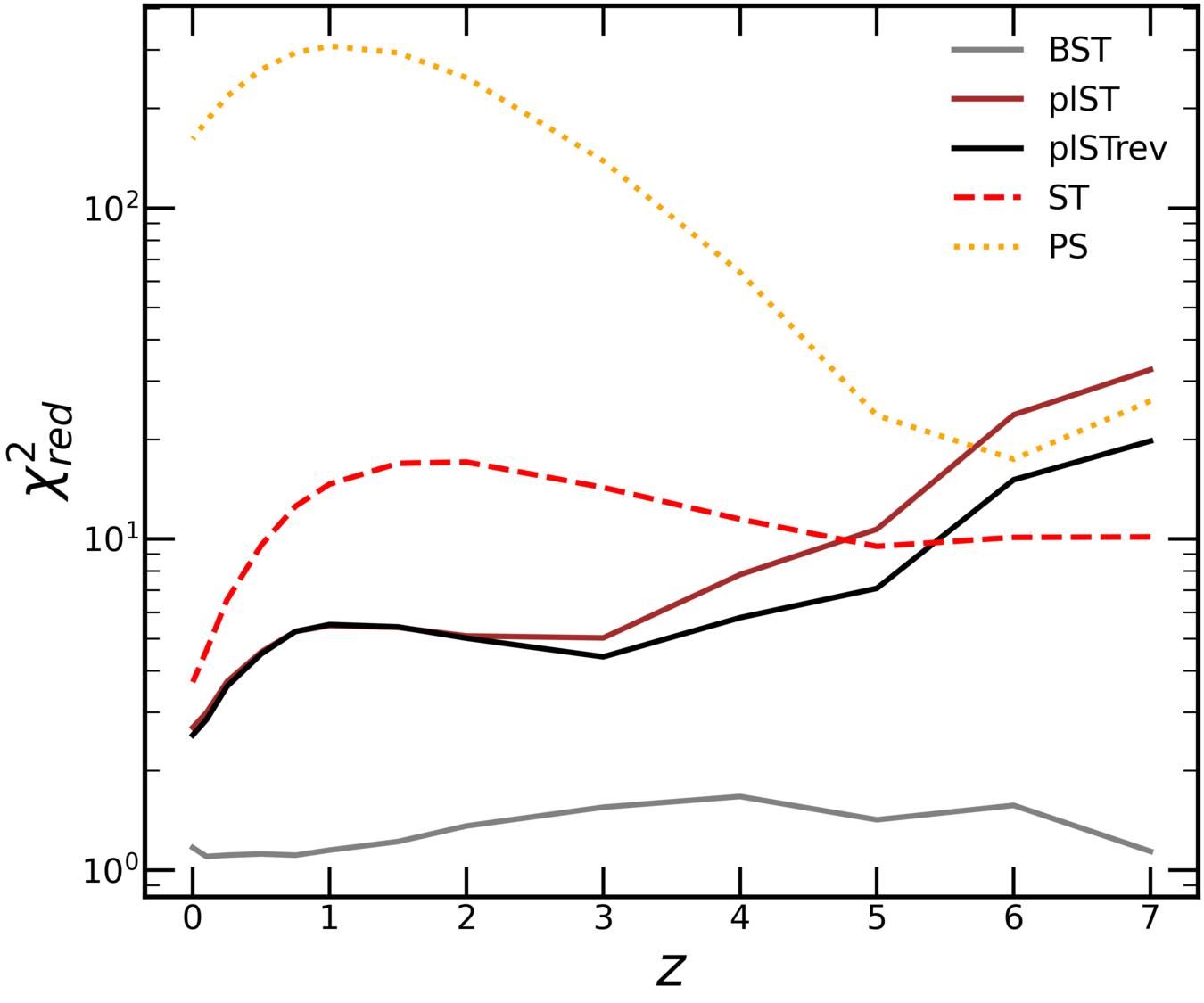}
    \caption{{\bf $\chi^2_{red}$ values for LCDM mass function fits:} Gray curve is $\chi^2_{red}$ for best fit Sheth-Tormen mass function, Brown and black curve are power law models prediction of with respect to equation \ref{eqn:pq_n} and \ref{eqn:pq_n_revised} respectively, dashed red curve is $\chi^2_{red}$ for standard Sheth-Tormen mass function, and the dotted orange curve is $\chi^2_{red}$ for Press-Schechter mass function.
    }
    \label{fig:lcdm_chi}
\end{figure}

In the last section, we have demonstrated conclusively that the mass functions for power law models in an EdS background are not universal. 
Departure from universality is small but non-zero.  
The set of models used to demonstrate this departure from universality is simple; indeed, it is the simplicity that allows us to demonstrate deviations from a universal behaviour.  
In this section, we seek to analyse how this departure from universality may affect the mass function in a more realistic model. 

Departures from universality in mass function can potentially arise from cosmology via the growth factor or the threshold for collapse, or it can come from the shape of the power spectrum.  
Our study has given us a quantitative handle on the variation in mass function with the slope of the power spectrum. 
In this section, we use the dependence of $p$ and $q$ on $n$ to estimate the expected variation of the mass function with the shape of the power spectrum.  
In order to map the varying slope of the LCDM power spectrum, we define an effective index at any redshift $z$ by computing
\newcommand\at[2]{\left.#1\right|_{#2}}
\begin{equation}
   \at{\frac{d \log \sigma(m,z)}{d \log m}}{\sigma=1} = - \frac{n_{eff}+3}{6}
    \label{eqn:sigma_d}
\end{equation}

Then we fit free ST to LCDM simulations epoch wise and compare the best fit p and q with equation \ref{eqn:pq_n}.
We use three Planck Cosmology ($\Omega_{m0}$= 0.3111, $\Omega_{\Lambda0}$ = 0.6889, $\Omega_{b0}$ = 0.0490, H0 = 67.66, $\sigma_8$ = 0.8102, $n_s$=1.0) \citep{Planck} simulations with box sizes 150, 300 and 500 Mpc with $1024^3$ particles. 
Small box simulations provide better mass resolution, which is essential to capture low mass halos. 
Further capturing low mass halos allows us to probe early epochs, as massive halos are rare.
On the other hand, large box simulations provide us with high mass halos.
We use one intermediate box to connect both mass ends.
We calculate mass functions using the $M_{200b}$ halo mass definition, 300 halos per bin, $N_{cut}=100$, and removal of the top 0.1\% of massive halos from each box. We verify that the mass function has converged with regard to box size by fitting individual boxes and comparing the results to a combined fit.
Figure \ref{fig:mfl} shows the residue of the mass function after fitting the Sheth-Tormen function to LCDM simulation data. Different panels are for different representative epochs. Dotted orange and dashed red lines represent the Press-Schechter and the Sheth-Tormen mass function, respectively. Dashed green lines represent equation \ref{eqn:pq_n_revised}. Solid black lines represent the best fit Sheth-Tormen mass function after $\chi^2$ analysis. Brown-gray data points show the mass function calculated from the simulations. Three different colors of data points represent different simulations used for analysis. Gray filled area shows the 1$\sigma$ confidence interval for the fit.

We show these fitted parameters in figure \ref{fig:lcdm_pq}, green data points show the best fit ST values of p and q for redshift.
Black data points represent predictions from power law models according to equation \ref{eqn:pq_n_revised}.
Brown scatter in the bottom panel shows equation \ref{eqn:pq_n} prediction.
Errorbars represent 1$\sigma$ errors.
Dashed red lines show p and q values from standard ST.
We find that the trend in p for both predicted and fitted values is similar. Their actual values are off by some factor, possibly because of cosmological contributions.
q in both cases is consistent within errorbars.
As we increase redshift, the halo count decreases, so challenging to probe that part.
It is not easy to probe mass function below $\nu<1$

In figure \ref{fig:lcdm_chi} we present $\chi^2_{red}$ values of LCDM mass function fit with epoch.
The gray curve shows $\chi^2_{red}$ for the best fit Sheth-Tormen mass function.
Brown and black curve represent power law models prediction of Sheth-Tormen mass function from equation \ref{eqn:pq_n} and \ref{eqn:pq_n_revised} respectively.
The dashed red and dotted orange curves show $\chi^2_{red}$ for standard Sheth-Tormen and Press-Schechter mass functions, respectively.

We find that power law prediction is a better fit to simulation data over standard ST for z<= 4.0.
For z>4.0, we have less number of data points resulting in large $\chi^2_{red}$ values.
Using revised fits for p and q (equation \ref{eqn:pq_n_revised}), $\chi^2_{red}$ for power law prediction improves further.
This indicates the scope of improvement in ST mass function.

\section{Conclusions}
\label{sec:conclusion}

We have used power law models in the Einstein-deSitter background to investigate the mass function of collapsed halos in cosmological N-Body simulations.
We fit the mass function obtained from simulations with the Sheth-Tormen form to obtain the parameters $p$ and $q$.  
This is done for a number of models with different power law indices in the range $-2.2 \leq n \leq 0$.
We summarise key findings of this work below.
\begin{itemize}
    \item 
    We find clear evidence that the halo mass function is not universal and has explicit power spectrum dependence. 
    \item 
    The Sheth-Tormen parameter values that are obtained by fitting the mass function show a systematic trend with the power-law index (see figure \ref{fig:mf_1} and \ref{fig:mf_2}. We also show a correlation between these parameters (see figure \ref{fig:pq_n}).
    \item 
    We find that the halo mass function has a weak epoch dependence: the best fit value of Sheth-Tormen parameters varies with redshift though the values are consistent with a constant value for a given power law model within 1$\sigma$ limit.
    \item
    We have shown that these results are not sensitive to the process used for constructing the halo catalog or subsequent analysis.
   \item    
    Best fit functions and the corresponding $\chi^2_{red}$ values indicate, as is also seen in the plots for residuals, that the Sheth-Tormen mass function form is inadequate and better modelling of collapse with more parameters may be required. 
    \item 
    We show that the Sheth-Tormen mass function with parameter values derived from a matched power-law EdS cosmology provides a better fit to the LCDM mass function than the standard Sheth-Tormen mass function. As we have not studied departures from universal mass function with cosmology, we cannot rule out a similar contribution from such dependence, especially at low redshifts.
\end{itemize}

While we have not studied the cause of non-universality in mass functions, we can speculate about its origin.  
The variation we have seen is with the slope of the power spectrum, and in these models, the only contribution that can cause this can arise from the coupling of modes between the collapsing structures and the large scale density field.  
While such coupling is not significant for spherical collapse, it becomes an essential ingredient for ellipsoidal collapse. 
Such coupling can be studied in N-Body simulations, and we propose to do so in our follow up work.

\section*{Acknowledgements}

Computational work for this study was carried out at the Saha and \href{https://www.niser.ac.in/hpc}{Kalinga} cluster in the National Institute of Science Education and Research (NISER), Bhubaneswar. NK is supported by the IUCAA\footnote{Inter-University Centre for Astronomy and Astrophysics, Pune, India} 
Associateship Programme. GK is partly supported by the Department of Atomic Energy (Government of India) research project with Project Identification Number RTI~4002, and by the Max Planck Society through a Max Planck Partner Group.

\section*{Data Availability}
The SUBFIND halo catalogs and GADGET4 snapshots of the simulations discussed in the text can be made available upon request.


\bibliographystyle{mnras}

\bsp	
\label{lastpage}
\end{document}